\UseRawInputEncoding
\pdfoutput=1
\documentclass[acmlarge,nonacm]{acmart}

\usepackage{algorithmic}
\usepackage{graphicx}
\usepackage{textcomp}
\usepackage{xcolor}
\usepackage{tcolorbox}

\usepackage{threeparttable}
\usepackage{multirow}
\usepackage{xspace}
\usepackage{booktabs}
\usepackage{rotating}
\usepackage{array}
\usepackage{makecell}
\usepackage{bbding}
\usepackage{longtable}
\usepackage{bbding}
\usepackage{pifont}
\usepackage{stfloats}

\AtBeginDocument{%
  }

\acmJournal{POMACS}
\acmVolume{37}
\acmNumber{4}
\acmArticle{111}
\acmMonth{8}

\begin{document}

\title{Sok: Comprehensive Security Overview, Challenges, and Future Directions of Voice-Controlled Systems }

\author{Haozhe Xu}
\email{haozhexu@whu.edu.cn}
\orcid{0009-0008-8264-725X}
\affiliation{%
  \institution{Wuhan University}
  \city{Wuhan}
  \state{Hubei}
  \country{CN}
}

\author{Cong Wu}
\orcid{0000-0002-0930-0283}
\affiliation{%
  \institution{Wuhan University}
  \city{Wuhan}
  \state{Hubei}
  \country{CN}
}
\email{cnacwu@whu.edu.cn}

\author{Yangyang Gu}
\email{guyangyang@whu.edu.cn}
\affiliation{%
  \institution{Wuhan University}
  \city{Wuhan}
  \state{Hubei}
  \country{CN}
}

\author{Xingcan Shang}
\email{xc_shang@whu.edu.cn}
\affiliation{%
  \institution{Wuhan University}
  \city{Wuhan}
  \state{Hubei}
  \country{CN}
}

\author{Jing Chen}
\email{chenjing@whu.edu.cn}
\orcid{0000-0002-7212-5297}
\affiliation{%
  \institution{Wuhan University}
  \city{Wuhan}
  \state{Hubei}
  \country{CN}
}

\author{Kun He}
\email{hekun@whu.edu.cn}
\orcid{0000-0003-3472-419X}
\affiliation{%
  \institution{Wuhan University}
  \city{Wuhan}
  \state{Hubei}
  \country{CN}
}

\author{Ruiying Du}
\email{duraying@whu.edu.cn}
\affiliation{%
  \institution{Wuhan University}
  \city{Wuhan}
  \state{Hubei}
  \country{CN}
}

\renewcommand{\shortauthors}{Xu et al.}

\begin{abstract}
The integration of Voice Control Systems (VCS) into smart devices and their growing presence in daily life accentuate the importance of their security. 
Current research has uncovered numerous vulnerabilities in VCS, presenting significant risks to user privacy and security. 
However, a cohesive and systematic examination of these vulnerabilities and the corresponding solutions is still absent. 
This lack of comprehensive analysis presents a challenge for VCS designers in fully understanding and mitigating the security issues within these systems.

Addressing this gap, our study introduces a hierarchical model structure for VCS, 
providing a novel lens for categorizing and analyzing existing literature in a systematic manner. 
We classify attacks based on their technical principles and thoroughly evaluate various attributes, 
such as their methods, targets, vectors, and behaviors. 
Furthermore, we consolidate and assess the defense mechanisms proposed in current research, 
offering actionable recommendations for enhancing VCS security. 
Our work makes a significant contribution by simplifying the complexity inherent in VCS security, 
aiding designers in effectively identifying and countering potential threats, and setting a foundation for future advancements in VCS security research.
\end{abstract}





\maketitle

\section{Introduction}
VCS, offering convenient user interfaces, have become integral to modern smart devices, including mobile technologies and smart home systems. The evolution of VCS has led to their widespread adoption, with the market projected to reach \$7.07 billion by 2023 \cite{Smart2023}. This surge in popularity is attributed to their diverse applications, ranging from managing smart homes to facilitating online shopping. However, this expansion brings significant security challenges, such as data privacy and vulnerability to cyber attacks, which have become increasingly critical concerns in the VCS domain.

VCS, as a complex amalgamation of software and hardware components, inherently possesses diverse security vulnerabilities. These vulnerabilities open avenues for attackers to devise a range of attack methods, challenging VCS designers to anticipate and understand these multifaceted threats. For example, some attackers exploit hardware vulnerabilities in microphones, using non-acoustic means like lasers to inject malicious commands into VCS \cite{sugawara2020light}. Furthermore, through adversarial training, attackers can create audio commands that, while seemingly benign to humans, are interpreted as malicious instructions by VCS. In the realm of third-party VCS services, attackers may embed carefully crafted malicious services, which could be inadvertently triggered by the VCS \cite{kumar2018skill, zhang2019dangerous}. However, most existing defense strategies focus narrowly on specific attack types, overlooking the broader implications of various attacks on the entire VCS ecosystem. This narrow focus leaves a gap in understanding the comprehensive risks and attack vectors facing VCS, beyond individual defense measures. Therefore, given these complexities, there is an urgent need for researchers in the VCS field to conduct a comprehensive and systematic survey, analyzing both the current attacks on VCS and the existing defense mechanisms.

\emph{Comparison with related surveys.} Our survey distinguishes itself from prior research by categorizing VCS into distinct layers: hardware, preprocessing, kernel, and service. This approach enables a comprehensive examination of security risks and defense mechanisms, considering each layer's specific integration perspective. Table \ref{tab:paper oberview} provides a comparative summary of our work in relation to similar studies. For instance, Zhang et al. \cite{zhang2019dangerous} focused specifically on security threats from malicious services within VCS. Bai et al. \cite{bai2020acoustic} explored acoustic technologies and their applications in domestic and industrial settings. Abdullah et al. \cite{abdullah2021sok} and Chen et al. \cite{chen2022sok} conducted in-depth analyses of risks associated with Automatic Speech Recognition (ASR) and Speaker Verification (SV) systems, integral components of VCS. Zhang et al. \cite{zhang2022adversarial} provided a detailed survey on adversarial attacks targeting ASR. These studies, while insightful, often focus on specific layers or subsets of layers within VCS, possibly overlooking the full view of security threats. For example, \cite{zhang2019dangerous} and \cite{bai2020acoustic} did not comprehensively address vulnerabilities in the service layer. Similarly, \cite{abdullah2021sok}, \cite{chen2022sok}, and \cite{zhang2022adversarial} primarily analyzed the kernel layer, neglecting other critical layers. In contrast, our survey aims to bridge these gaps by providing an all-encompassing analysis of VCS security across all layers. Although the work in \cite{edu2020smart} presents a comprehensive overview of VCS attacks, it predominantly focuses on smart home applications, omitting extensive discussion on mobile device integration.

\emph{Motivation.}
In addition to the current surveys lacking a comprehensive perspective on examining VCS security issues, researchers also urgently need to address the following questions:
\begin{itemize}
    \item VCS systems are vulnerable to a variety of attacks, such as transduction attacks, speech synthesis attacks, and adversarial attacks. This susceptibility stems from two main factors: the advancement of technology providing attackers with more sophisticated tools and algorithms, and the inherent complexity of VCS as a system comprising multiple hardware and software components, each with unique security vulnerabilities. Understanding the mechanics behind these attacks and the specific vulnerabilities of VCS components is crucial for designers.
    \item Beyond understanding attack mechanisms, it is vital for VCS designers to be aware of effective defense solutions. Current defense strategies, such as adversarial training, are often tailored to specific attack types \cite{wang2021adversarial, gong2018overview, Iter2017GeneratingAE, yakura2018robust, abdel2014convolutional, hai2003improved, zeng2019multiversion}, posing a challenge in selecting appropriate defense combinations to robustly enhance VCS against a wide array of attacks.
    \item The diverse nature of VCS, encompassing both automatic speech recognition (ASR) and speaker verification (SV), often leads to these components being studied in isolation \cite{abdullah2021sok, chen2022sok}. However, many VCS systems incorporate both ASR and SV functionalities, and attacks against these systems often exhibit overlapping characteristics. Designers must, therefore, consider the potential vulnerabilities of both ASR and SV in tandem when designing VCS.
\end{itemize}

\begin{figure*}[!t]
  \centering
  \includegraphics[width=.76\textwidth, bb = 0 0 1311 1609]{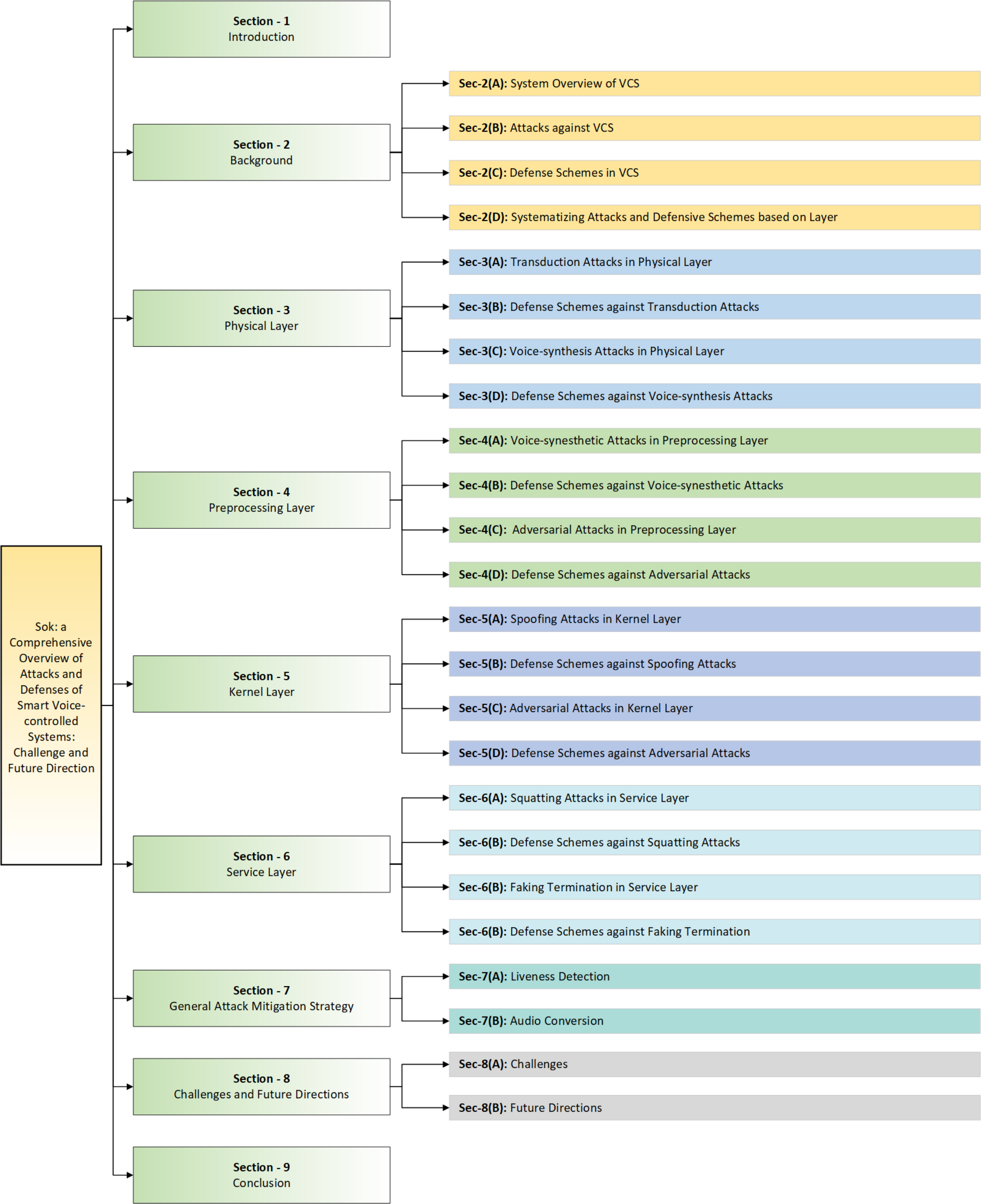}
  \caption{An Illustrative Overview of Structure of Our Survey Article.}
  \label{fig: Paper_structure}
  \Description[paper structure]{Divide the paper into 8 sections.}
\end{figure*}

In our paper, we conduct a comprehensive and systematic study of VCS security. 
To thoroughly discuss the various attacks VCS faces, we categorize these attacks based on the hierarchical model of VCS, which includes hardware, software, and service layers. We then analyze and assess the corresponding defense schemes, tailored to each type of attack within these layers. For each attack, we conduct a multi-dimensional evaluation, considering aspects such as attack methods, targets, vectors, and performance impact, and discuss potential directions for future research. Additionally, we offer practical combinations of defense schemes, providing valuable insights for VCS researchers to effectively counter diverse attacks. An illustrative organizational structure for our survey article is presented in Fig. \ref{fig: Paper_structure}.

Our work distinguishes itself from existing surveys in several comprehensive aspects: i) We provide extensive coverage of a broad range of attack methods applicable to each layer within the hierarchical model of VCS. This approach offers a nuanced understanding of the potential vulnerabilities at different levels of the system. ii) Our overview is expanded to include various devices that incorporate VCS, where we detail the unique security risks associated with each device type. This broadens the scope of our survey to encompass a wider array of real-world applications. iii) We present a detailed introduction to defense schemes corresponding to these attack methods, thereby enabling readers to not only understand these strategies but also to implement them in practical scenarios. This comprehensive approach to both attack and defense aspects provides a well-rounded perspective on VCS security.

\begin{table*}[!t]
  \scriptsize
  \renewcommand\arraystretch{1.3}
  \centering
  \caption{A comparison of existing survey work in the smart voice-controlled systems}
  \label{tab:paper oberview}
  \begin{threeparttable}
    \begin{tabular}{ccp{20em}cccccccccc}
    \toprule
    \multirow{3}{*}{Ref No.} & \multirow{3}{*}{Year} & \multicolumn{1}{c}{\multirow{3}{*}{Major Contribution of Surveys}} & \multicolumn{10}{c}{Scope} \\
\cline{4-13}    \multicolumn{1}{c}{} &       & \multicolumn{1}{c}{} & \multicolumn{5}{c|}{Attack}   & \multicolumn{5}{c}{Defense} \\
\cline{4-13}    \multicolumn{1}{c}{} &       & \multicolumn{1}{c}{} & \multicolumn{1}{c}{SVS\tnote{1}} & \multicolumn{1}{c}{NLPS\tnote{2}} & \multicolumn{1}{c}{HS\tnote{3}} & \multicolumn{1}{c}{SWSS\tnote{4}} & \multicolumn{1}{c|}{MDS\tnote{5}} & \multicolumn{1}{c}{SVS} & \multicolumn{1}{c}{NLPS} & \multicolumn{1}{c}{HS} & \multicolumn{1}{c}{SWSS}  & \multicolumn{1}{c}{MDS}\\
    \midrule
    \rule{0pt}{16pt} 
    \cite{zhang2019dangerous} & 2019  & A comprehensive study about third party skills with threats in VA and how to avoid them. & \XSolidBrush     & \XSolidBrush     & \XSolidBrush     & \Checkmark    & \Checkmark & \XSolidBrush     & \XSolidBrush     & \XSolidBrush     & \Checkmark & \Checkmark\\
    \rule{0pt}{16pt}
    \cite{edu2020smart} & 2020  & A thorough survey over the security and privacy of Personal Assistant of smart home personal assistants & \Checkmark     & \Checkmark     & \Checkmark     & \Checkmark     & \XSolidBrush & \Checkmark     & \Checkmark     & \Checkmark     & \Checkmark & \XSolidBrush\\
    \rule{0pt}{16pt}
    \cite{bai2020acoustic} & 2020  & A comprehensive survey on acoustic-based technologies and applications in life and industry. & \Checkmark     & \XSolidBrush     & \Checkmark     & \XSolidBrush     & \Checkmark & \XSolidBrush     & \XSolidBrush     & \XSolidBrush     & \XSolidBrush & \Checkmark\\
    \rule{0pt}{16pt}
    \cite{abdullah2021sok} & 2021  & A detailed study about security threats and defense schemes in ASR and SV. & \Checkmark     & \XSolidBrush     & \XSolidBrush     & \XSolidBrush     & \Checkmark & \Checkmark     & \XSolidBrush     & \XSolidBrush     & \XSolidBrush & \Checkmark\\
    \rule{0pt}{16pt}
    \cite{chen2022sok} & 2022  & A modularized and comprehensive survey about ASR security and comparison between ASR and IRS security. & \XSolidBrush     & \Checkmark     & \XSolidBrush     & \XSolidBrush     & \Checkmark & \XSolidBrush     & \Checkmark     & \XSolidBrush     & \XSolidBrush & \Checkmark\\
    \rule{0pt}{16pt}
    \cite{zhang2022adversarial} & 2022  & A detailed investigation of adversarial attacks against ASR. & \XSolidBrush     & \XSolidBrush     & \XSolidBrush     & \XSolidBrush     & \Checkmark & \XSolidBrush     & \XSolidBrush     & \XSolidBrush     & \XSolidBrush & \Checkmark\\
    \rule{0pt}{16pt}
    \makecell[c]{Ours} & 2023  & A comprehensive overview of the challenges faced by VCS and its future direction. & \Checkmark     & \Checkmark & \Checkmark     & \Checkmark     & \Checkmark     & \Checkmark     & \Checkmark     & \Checkmark     & \Checkmark & \Checkmark\\
    \bottomrule
    \end{tabular}%
    \begin{tablenotes}
      \footnotesize
      \item $^{1}$SVS: Speaker Verification Security. 
      $^{2}$NLPS: Natural Language Processing Security. 
      $^{3}$HS: Hardware Security.
      $^{4}$SWSS: Skill Web Service Security.  
      $^{5}$MDS: Mobile Device Security.
    \end{tablenotes} 
  \end{threeparttable} 
\end{table*}%

\emph{Contributions.}
We summarize our contribution as follows:
\begin{itemize}
    \item Our unique approach to examining VCS security involves dividing VCS into four distinct layers: the physical layer, preprocessing layer, kernel layer, and service layer. This hierarchical model allows for a systematic and detailed analysis of both attacks and defense schemes at each layer, providing VCS researchers with a comprehensive framework to understand and evaluate associated risks accurately.
    
    \item Utilizing our proposed hierarchical model, we have systematically integrated and assessed potential attacks on each VCS layer. By employing various metrics, such as attack complexity and potential impact, we offer a feasibility analysis that enables users to gauge the security threats these attacks pose to VCS accurately.

    \item We have developed and evaluated defense strategies tailored to each layer of VCS, considering practical effectiveness and usability from multiple dimensions. Additionally, we introduce a generalized attack mitigation strategy, aiding designers in constructing a more robust and comprehensive defense system for VCS.

    \item Through an extensive analysis of existing literature, we provide targeted recommendations for enhancing current security research and directions for future development in the field, addressing gaps and emerging trends in VCS security research.
\end{itemize}

\section{Background}\
In this section, we brief the common system overview of VCS, then illustrate attacks and defenses of VCS.
It then systematize attacks and defensive schemes based on Layer
\subsection{System Overview of VCS}

\begin{figure*}[ht]
  \centering
  \includegraphics[width=.9\textwidth, bb = 0 0 2304 293]{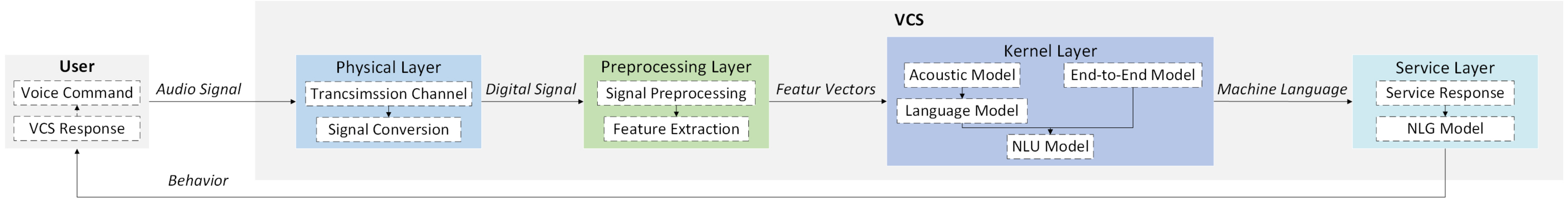}
  \caption{A hierarchical Workflow Diagram of VCS.  After a user issues a voice command to the VCS, the user will receive a response from the VCS. Based on the response, the user can choose to issue another relevant command to the VCS.}
  \Description[workflow]{Use five block diagrams to represent the structure of VCS.}
  \label{fig: VCS_construction} 
\end{figure*}

As Siri achieved significant success on iPhones, VCS have seen widespread integration into daily life. Nowadays, most smart devices, including smartphones, smart homes, and smart vehicles, come equipped with voice assistants. VCS enable users to perform various tasks such as unlocking devices, setting alarms, and opening doors, thereby enhancing interaction with smart technology. Popular VCS like Google Home, Amazon Echo, and Apple HomePod have expanded their capabilities through third-party services, providing features like voice translation and audio transcription.

With the rapid development and increasing prevalence of VCS, security concerns have similarly grown. To thoroughly analyze and effectively organize the potential risks and defensive strategies for voice assistants, we categorize VCS into four functional layers: the physical layer, preprocessing layer, kernel layer, and service layer. This layered approach allows for a detailed examination of how VCS process voice commands and provide services to users, as illustrated in Fig.~\ref{fig: VCS_construction}.

\subsubsection{Physical Layer}
\emph{Transmission channels.}
Acoustic signals, when transmitted over-the-air in VCS, are subject to various environmental influences that can degrade their quality. Ideally, signal transmission should occur without loss, but realistically, signal intensity diminishes with distance, reducing the information carried. This limitation poses a challenge in capturing sufficient signal data, thus constraining the effective operational range between the source and the target device. Environmental factors such as noise types can also hinder a VCS's ability to accurately interpret acoustic signals. Furthermore, signal reflection off surrounding objects can result in the reception of mixed signals, complicating the interpretation of the original acoustic input.

This vulnerability is particularly concerning in security contexts, where attackers could exploit these transmission challenges by deploying malicious audio. Such attacks might aim to compromise the VCS by leveraging signal interference, necessitating countermeasures. Attackers, aware of these transmission challenges, might enhance the robustness of their malicious audio or resort to generating electronic signals that mimic acoustic ones, to increase the likelihood of successful system penetration.

\emph{Acoustic signal conversion.}
For a VCS to process acoustic signals, they must first be converted into digital signals, a process involving two critical steps. Initially, the system employs sensors, typically microphones, to transform acoustic signals into electrical signals. Among microphones, electrodynamics and capacitive types are common, with the latter being more prevalent in modern devices due to their sensitivity and reliability. Capacitive microphones operate by detecting changes in capacitance caused by the vibration of movable plates in response to sound waves, thereby converting acoustic signals into electrical ones \cite{zawawi2020review}.

The next stage involves an analog-to-digital converter (ADC), which digitizes the electrical signals. The ADC employs a sample and hold circuit to capture and stabilize the signal value, followed by a quantizer that processes these samples, converting them into binary numbers representing the digital signal. This conversion is pivotal for effective VCS processing, but each stage also presents potential security vulnerabilities, such as susceptibility to signal manipulation or interference, that must be carefully managed to ensure system integrity.

\subsubsection{Preprocessing Layer}
\emph{Signal preprocessing.}
Signal preprocessing is vital in VCS, as it involves processing raw speech signals to extract clean speech for analysis. VCS utilize band filters to isolate speech signals within the 300-3400kHz frequency band, essential for clear speech recognition. The use of multiple microphones for multichannel speech capture necessitates techniques such as multichannel acoustic echo cancellation, reverberation control, and source separation. These processes ensure high-quality signal capture, crucial for accurate speech recognition. Recent advancements include the application of deep learning methods for signal preprocessing, offering improved noise mitigation and signal clarity~\cite{ryant2013speech}. This enhanced preprocessing is critical for robust VCS functioning, particularly in complex acoustic environments.

\emph{Acoustic feature extraction.}
This step aims to extract distinctive features from speech signals, accounting for variations in speaker characteristics, environmental noise, and transmission channels. The process begins with segmenting the processed audio into frames using a sliding window technique, followed by feature extraction. Common algorithms for feature extraction include Short-time Fourier Transform (STFT), Linear Predictive Coding (LPC)~\cite{rabiner1993fundamentals}, Mel Frequency Cepstral Coefficients (MFCC)~\cite{sigurdsson2006mel}, and Perceptual Linear Prediction (PLP)~\cite{rabiner1978digital}. Each algorithm plays a unique role in enhancing the robustness of VCS by efficiently capturing and representing different acoustic properties of speech.

\subsubsection{Kernel Layer}
\emph{Acoustic model.}
Post feature extraction in the preprocessing layer, the acoustic model establishes a statistical relationship between these features and acoustic units like phonemes using machine learning algorithms. Traditional models commonly use Gaussian Mixture Models (GMM) and Hidden Markov Models (HMM), while contemporary models increasingly employ neural networks such as Convolutional Neural Networks (CNN) and Recurrent Neural Networks (RNN) \cite{sainath2015convolutional}, enhancing the accuracy and efficiency of phoneme conversion.

\emph{Language model.}
The language model predicts the likelihood of word sequences forming coherent sentences, differentiating between homophones based on context. This model is pivotal in controlling VCS output using common words and grammatical rules. Models are generally divided into statistical (e.g., N-gram, HMM) and neural network-based, with the latter becoming increasingly prevalent in modern VCS.

\emph{Phonetic dictionary.}
This component links the acoustic and language models by mapping words to phonemes, essential for decoding the user's intent in VCS. The phonetic dictionary facilitates the establishment of relationships between the models' units, thereby aiding in effective system decoding.

\emph{End-to-end model.}
To streamline VCS, end-to-end models have been developed. These models directly convert acoustic features into word sequences, integrating the functions of both acoustic and language models. This simplification reduces potential attack vectors, enhancing system security. Examples include CTC \cite{graves2006connectionist}, Transducer \cite{graves2012sequence}, and Attention models \cite{chorowski2015attention}.

\emph{Natural language processing (NLP).}
NLP enables machines to understand and act upon natural language inputs. Its development has evolved from rule-based to statistical models, and now to neural network-based models like Word2vec \cite{mikolov2013efficient} and BERT \cite{devlin2018bert}, reflecting the industry's shift towards more sophisticated, context-aware processing.

\emph{Speaker verification.}
Distinct from traditional VCS, speaker verification focuses on identifying the speaker rather than interpreting speech content. It consists of speaker identification (SI) and speaker verification (SV), with the latter confirming a speaker's identity. In the kernel layer, speaker verification differs from standard VCS by determining the probability of potential speakers, thus requiring specialized discussion regarding its security implications.

\subsubsection{Service Layer}
After interpreting the user's intent in the kernel layer, the VCS responds in the service layer. Utilizing Natural Language Generation (NLG), the system conveys the required information in text or audibly through Text-to-Speech (TTS) technology. Technological advancements have expanded VCS capabilities to include third-party services, enabling more complex functionalities. However, this expansion introduces new security challenges, such as increased exposure to data breaches and exploitation of service vulnerabilities, necessitating robust security measures.

\subsection{Attacks against VCS}
Current researches highlight the vulnerability of VCS to a range of attacks, stemming from its complex architecture which presents potential vulnerabilities at each layer. To equip system designers with a comprehensive understanding of these threats, this section delves into the nature of attacks against VCS, examining them from the perspectives of threat models, attack metrics, and attack objectives. We will explore how threat models help identify potential security weaknesses, discuss metrics for assessing the impact and severity of attacks, and analyze the typical goals of attackers targeting VCS systems.


\subsubsection{Threat Model}
We consider attack scenarios under these assumptions:
\begin{enumerate}
  \item Attackers may have prior knowledge of the target VCS, including its brand and model. For common VCS, attackers might possess the same device model.
  \item The VCS operates on a trusted platform with encrypted communication, preventing attacks via the operating system, hardware, or communication interception.
  \item Attackers' knowledge varies: 
  1) White/Grey-box: Full or partial model knowledge, applicable to open-source VCS.
  2) Black-box: No model knowledge, typical in commercial VCS.
\end{enumerate}

\subsubsection{Attack Metrics}
Based on the threat model, we assess VCS attacks as follows:

\emph{Attack method.} The method determines the targeted VCS layer. Some exploit non-audio signals to attack the physical layer, while others use audio-based methods to deceive the kernel layer.

\emph{Attack target.} Targets may vary from specific models (requiring detailed knowledge) to general hardware vulnerabilities affecting multiple VCS models.

\emph{Attack vector.} Over-the-air attacks, more feasible in real-world settings, must be robust against environmental factors and signal degradation.

\emph{Performance.} The effectiveness of an attack is measured by its success rate and maximum effective distance, indicating its practicality in real scenarios.

\subsubsection{Attacker Goal}
Attackers primarily aim to disrupt VCS operations, leading to unauthorized command execution. Untargeted attacks may cause incorrect audio decoding, while targeted attacks seek to execute specific malicious commands. Security impacts include unauthorized control of smart devices (e.g., unlocking doors), while privacy concerns involve eavesdropping or accessing personal information via malicious services.

\subsection{Defense Schemes in VCS}
The range of defense strategies for VCS is relatively narrow compared to the diversity of attack methods. Current defenses primarily fall into two categories: 1) Layer-specific solutions, such as adversarial training, which fortify particular layers like the kernel layer against specific types of attacks, yet they might not be effective against threats targeting other layers. This category also includes specialized encryption for communication security and hardware-focused safeguards. 2) Holistic strategies, like liveness detection, are designed to protect the entire VCS by verifying the authenticity of voice inputs, thereby securing all layers from a range of potential attacks. While liveness detection is adept at confirming genuine user interaction, challenges such as differentiating between live and recorded voices persist. The effectiveness of VCS security often hinges on integrating both layer-specific and holistic defenses to comprehensively address the multifaceted nature of potential threats.

\subsection{Layer-Based Systematization of Attacks and Defenses in VCS}

The prevailing approach in academic circles for classifying attacks on VCS typically revolves around either human perception or specific VCS components. However, this method can inadvertently group distinct attack methodologies under the same category, despite their differing technical principles. For instance, both DolphinAttack~\cite{zhang2017dolphinattack} and Hidden Voice Command~\cite{carlini2016hidden} are categorized as voice-synthesis attacks. Nonetheless, they fundamentally differ in their mechanisms: DolphinAttack exploits microphone hardware vulnerabilities, whereas Hidden Voice Command targets the preprocessing algorithms. 

While the component-based classification method provides a systematic approach to categorizing attacks, it often leads to an isolated analysis of SV and ASR components, as seen in existing literature~\cite{chen2022sok, abdullah2021sok}. Our research suggests that despite the distinct outputs of these two components, they share similar technical principles and vulnerabilities that can be exploited by attackers. Thus, analyzing them separately may obscure the full understanding of potential attack vectors.

To address these classification challenges, our article proposes a restructuring of attack categorization based on the VCS architecture, dividing them into attacks on the physical layer, preprocessing layer, core layer, and service layer, as outlined in Fig. \ref{fig: VCS_construction}. This approach allows for a more precise identification of the vulnerabilities specific to each layer and the attacks that exploit them. We will also categorize defense solutions accordingly: those designed to protect specific layers will be discussed in their respective sections, while generalized defense strategies applicable across all VCS layers will be comprehensively covered in Section \ref{sec:GDS}. The ensuing chapters will detail the array of attack threats and corresponding defensive measures for each layer within VCS.


\section{Physical Layer}
The physical layer of VCS plays a pivotal role in converting sound signals into digital formats. This conversion process involves the transmission of sound signals through channels, typically air, to microphone devices, which are then digitized. However, this layer is vulnerable to a range of attacks due to the lack of signal verification in the transmission channel and the susceptibility of microphone devices to manipulation. Attackers can exploit these vulnerabilities by transmitting malicious audios or using specific signals to coerce microphones into generating attacker-desired digital signals. This section explores various attack types targeting the physical layer, including transduction attacks that manipulate physical components to create false inputs, voice-synthesis attacks that craft misleading audio inputs, and denial of service attacks aimed at overwhelming the system. A comprehensive summary and comparison of these attacks are detailed in Table~\ref{tab:physical layer}, elucidating their methods, targets, and potential impacts on the VCS.

\subsection{Transduction Attacks in Physical Layer}
Transduction attacks exploit microphone vulnerabilities in VCS by using non-standard signals that microphones should typically not recognize. These attacks manipulate the physical properties of microphone equipment to induce unintended digital signal generation, leading to VCS compromise.

\emph{Light Signal.} Microphones, designed to be sensitive only to sound signals, can unexpectedly respond to modulated light signals. Sugawara et al.~\cite{sugawara2020light} demonstrated that by modulating laser amplitude, attackers can transmit malicious commands to VCS from distances exceeding 100 meters, exploiting this unintended sensitivity.

\emph{Electromagnetic Wave (EW).} Microphone circuits can inadvertently act as antennas, receiving electromagnetic waves which are then demodulated into malicious commands. This vulnerability allows attackers to use electromagnetic coupling, as shown by Kasmi and Esteves~\cite{kasmi2015iemi}, to control voice assistants through headphone cables. Further research indicates that specific electromagnetic waves generated by wired~\cite{wang2022ghosttalk} or wireless chargers~\cite{dai2022inducing} can also inject commands into VCS.

\emph{Defense Schemes against Transduction Attacks.} For light-based attacks, sensor fusion technology~\cite{davidson2016controlling} offers a software-level defense by exploiting the omnidirectional nature of sound versus the directional nature of light, which typically affects only one microphone. This approach detects discrepancies between multiple microphone inputs. Hardware solutions include shielding microphones with opaque covers to block light, though this may attenuate sound signals~\cite{wang2015era}. Against EW attacks, sensors detecting electromagnetic changes can alert to potential intrusions. A comprehensive defense strategy involves adding authentication mechanisms to VCS, effectively mitigating various transduction attacks.

\subsection{Voice-synthesis Attacks in Physical Layer}
Voice-synthesis attacks in the physical layer of VCS differ from transduction attacks as they primarily utilize ultrasonic signals, which are inaudible to humans but can be received by most microphones. These attacks exploit the fact that ultrasound, with frequencies above 20 Hz, can be demodulated in the system through nonlinear amplifiers after being received by the microphone. Attackers modulate malicious commands into these ultrasonic signals, leading to interference with the normal functioning of VCS. Researchers have demonstrated such attacks using solid surface transmission\cite{yan2020surfingattack} and air transmission methods\cite{xia2023near, zhang2017dolphinattack, iijima2018audio, ji2021capspeaker, roy2018inaudible, song2017poster, yan2019feasibility}. Additionally, a novel approach exploiting specific current fluctuations to remotely attack VCS through sounds generated by the switching mode power supply (SMPS) in the same power grid was proposed in \cite{lanqing2023remote}.

\emph{Defense Schemes against Voice-synthesis Attacks.}
To counteract these attacks, hardware-based defenses include using microphones that are incapable of receiving ultrasonic signals, similar to the iPhone 6 Plus microphone\cite{xia2023near}, or implementing a "guarding" mechanism with an external signal generator to neutralize malicious audio\cite{he2019canceling}. However, replacing microphones or adding signal generators in existing VCS products is impractical, making software-level defenses more viable. For air-transmitted voice-synthesis attacks, the demodulated attack signal in the 500 to 1000 Hz range differs significantly from the original signal, detectable via a variational auto-encoder\cite{li2023learning}. Additionally, a microphone array can calculate the attenuation rate of sound waves at different microphones, identifying ultrasonic-origin voice commands\cite{zhang2021eararray}. Solid surface attacks lack this attenuation distinction, but their recovered attack signal introduces new frequency components from 10 kHz to 20 kHz, providing another basis for detection and differentiation from baseband signals.

\begin{table*}[!t]
    \scriptsize
    \renewcommand\arraystretch{1.7}
    \centering
    \caption{Comparison of transduction attack and voice-synthesis attack in physical layer}
      \begin{tabular}{c|c|c|c|c|c|c}
      \hline
      \multirow{2}{*}{Attack Type} & \multicolumn{1}{c|}{\multirow{2}{*}{Paper}} & \multicolumn{1}{c|}{\multirow{2}{*}{Attack Method}} & \multicolumn{1}{c|}{\multirow{2}{*}{Attack Target}} & \multicolumn{1}{c|}{\multirow{2}{*}{Attacker Vector}} & \multicolumn{2}{c}{Performance} \\
  \cline{6-7}          &       &       &       &       & Success & Distance \\
      \hline
      \multicolumn{1}{c|}{\multirow{4}{*}{\begin{sideways} \makecell[c]{Transduction \\ Attacks}\end{sideways}}} & \cite{sugawara2020light} & Light & Microphone & Over-the-air & 90\%  & 25m \\
  \cline{2-7}          & \cite{kasmi2015iemi} & EW    & Headphone Cable & Over-the-air & n.a     & n.a \\
  \cline{2-7}          & \cite{wang2022ghosttalk} & EW    & Charging Port & Power Line & 100\% & n.a \\
  \cline{2-7}          & \cite{dai2022inducing} & EW    & Microphone & Over-the-air & 91\%  & 5 cm \\
      \hline
      \multicolumn{1}{c|}{\multirow{9}{*}{\begin{sideways}Voice-synthesis Attacks\end{sideways}}} & \cite{yan2020surfingattack} & Ultrasonic & Microphone & Solid Surface & 100\% & 2.5m+ \\
  \cline{2-7}          & \cite{zhang2017dolphinattack} & Ultrasonic & Microphone & Over-the-air & 100\% & 165 cm \\
  \cline{2-7}          & \cite{iijima2018audio} & Ultrasonic & Microphone & Over-the-air & 100\% & 10 m \\
  \cline{2-7}          & \cite{ji2021capspeaker} & Ultrasonic & Capacitor & Over-the-air & 80\%  & 10.5 cm \\
  \cline{2-7}          & \cite{roy2018inaudible} & Ultrasonic & Microphone & Over-the-air & 50\%  & 760 cm \\
  \cline{2-7}          & \cite{song2017poster} & Ultrasonic & Microphone & Over-the-air & 80\%  & 3 m \\
  \cline{2-7}          &\cite{yan2019feasibility} & Ultrasonic & Microphone & Over-the-air & 100\% & 165 cm \\
  \cline{2-7}          &\cite{xia2023near}     & Ultrasonic & Microphone & Over-the-air & 100\% & 360 cm \\
  \cline{2-7}          &\cite{lanqing2023remote} & Audible Sound & SMPS & Over-the-air & 90\% & 23 m \\
      \hline
      \end{tabular}%
    \label{tab:physical layer}%
  \end{table*}%
  
\section{Preprocessing Layer}
In VCS, the preprocessing layer is essential for refining the digital signals obtained from the physical layer. This layer's primary function is to filter and enhance audio signals, which often contain a mix of human voices and background noise. Voice activity detection algorithms, like G.729~\cite{sohn1999statistical}, are used to isolate segments containing human speech, followed by the application of low-pass filters to reduce noise and improve VCS's recognition capabilities. Advances in Deep Neural Networks (DNN) have further enhanced these preprocessing tasks~\cite{ryant2013speech}. Besides noise reduction, this layer is responsible for extracting relevant features from audio signals using techniques such as DFC, MFCC, LPC, MFCS, and PLP, preparing them for subsequent phoneme, character, or word sequence recognition. However, vulnerabilities within these feature extraction processes have been identified, allowing attackers to execute voice-synthesis and adversarial attacks that deceive the VCS. This section discusses these attack methods, exploring how they exploit the preprocessing layer's weaknesses. Table \ref{tab:preprocessing layer} provides a comparative summary of these attacks, outlining their strategies and impact on VCS.

\subsection{Voice-synthesis Attacks in Preprocessing Layer}
Voice-synthesis attacks in the preprocessing layer of VCS utilize technology to create specific speech patterns that can be misinterpreted as legitimate instructions. Due to the lossy nature of signal processing and feature extraction in this layer, attackers can craft voice segments that mimic target voice features. These segments are then mistakenly recognized as intended voice signals by the VCS after undergoing preprocessing. A key characteristic of these attacks is the significant difference between the synthesized and target audio signals, making detection challenging for users.

For instance, Bispham et al.~\cite{bispham2019nonsense} produced nonsensical voice commands by altering consonant phonemes in the target voice. Despite the apparent errors, the VCS's voice model could correct these anomalies, leading to system recognition of these commands. Abdullah et al.~\cite{abdullah2019practical} employed various signal processing techniques like Time Domain Inversion and High Frequency Addition to morph the target voice into an unintelligible audio signal. They leveraged the VCS's sensitivity to specific keywords to construct more effective malicious samples. In another approach, Carlini and Vaidya et al.~\cite{carlini2016hidden, vaidya2015cocaine} targeted the Mel Frequency Cepstral Coefficients (MFCC) feature extraction, common in VCS. By calculating the MFCC features of the target voice and then reversing this process, they generated synthetic audio that successfully deceived the VCS. While these attacks effectively disguise their intent, the unusual noises they produce may still raise user suspicion.

\emph{Defense Schemes against Voice-synthesis Attacks.}
Defending against voice-synthesis attacks in the preprocessing layer of VCS involves two main strategies: detection and prevention. Detection refers to the system's ability to recognize an attack and alert the user. One common approach is the use of verification mechanisms like CAPTCHAs~\cite{matt2005inaccessibility}, where the system requests a verification response from the user before executing a voice command. This allows users to confirm if the command aligns with their intentions. Additionally, prompt warnings can notify users when the VCS receives voice commands, providing an opportunity to recognize and counteract potential attacks, even if the user may overlook the unusual noise generated by the attack~\cite{carlini2016hidden}.

Prevention strategies focus on averting the damage from an attack. Voice Activity Detection (VAD) is proposed as a preventive measure~\cite{abdullah2019practical}. As a speech processing algorithm, VAD can distinguish between noise and actual speech~\cite{hartpence2013packet, ramirez2007voice}, effectively identifying and disregarding synthesized audio attacks. Moreover, adjusting VCS filters to slightly degrade audio quality can disrupt synthetic audio, which typically lacks robustness, while still preserving the recognition of genuine human speech~\cite{carlini2016hidden}. These preventive measures work by exploiting the inherent weaknesses in synthesized audio, thereby reducing the likelihood of successful attacks.

\subsection{Adversarial Attacks in Preprocessing Layer}
Originally emerging in the field of image recognition~\cite{goodfellow2014explaining}, the concept of adversarial attacks has since gained significant attention in speech recognition. Contrary to voice-synthesis attacks, adversarial attacks craft voice commands that sound normal to human users but lead to incorrect predictions by the machine learning models in VCS. These attacks typically involve introducing subtle perturbations into benign audio samples, creating what are known as adversarial samples. These samples exploit the sensitivity of deep learning models to small input changes, posing serious security and privacy risks to audio-based systems.

In audio applications, these perturbations can be added directly to the audio waveform or to its features. When perturbations are applied to audio features, attackers must reverse-engineer the audio waveform from these features to ensure the altered audio is still recognizable by the VCS. This subsection focuses on adversarial attacks that perturb audio features, as this method intricately involves the operational principles of the preprocessing layer in VCS. Such attacks highlight the inherent vulnerabilities in the feature extraction and processing mechanisms, underscoring the need for robust defense strategies in this layer of VCS.

\subsubsection{Taxonomy of Adversarial Attacks in VCS}
Adversarial attacks in VCS can be classified based on different criteria, including the attack scenario, target, and vector.

\textbf{Attack Scenario:} 
Adversarial attacks are divided into white-box, black-box, and gray-box attacks based on the attacker's knowledge of the target VCS model. White-box attacks assume complete access to the VCS model, including its internal parameters. Black-box attacks, in contrast, limit the attacker to accessing only the inputs and outputs of the VCS, without knowledge of the internal model details. Gray-box attacks represent an intermediate scenario, where the attacker has partial knowledge of the model's parameters.

\textbf{Attack Target:} 
In terms of objectives, adversarial attacks can be either targeted or untargeted. Targeted attacks manipulate the VCS to produce a specific, incorrect output defined by the attacker. In contrast, untargeted attacks aim to induce any error in the system's output without a predetermined result. For example, in Speaker Verification (SV), untargeted attacks might result in the identification of any incorrect speaker, while targeted attacks would misidentify a speaker as a specific individual chosen by the attacker. Similarly, in Automatic Speech Recognition (ASR), untargeted attacks cause any incorrect transcription, whereas targeted attacks aim for a specific false transcription.

\textbf{Attack Vector:} 
Based on the method of delivery, adversarial attacks in VCS can be categorized into Wav-to-API (WTA) attacks and Wav-air-API (WAA) attacks~\cite{yuan2018commandersong}. WTA attacks involve directly inputting an adversarial audio file into the VCS API, whereas WAA attacks transmit the adversarial sample through the air, capturing it with the system’s microphone before processing. Additionally, there is the Wav-Wave-API (WWA) vector, where adversarial audio deceives VCS after transmission through a transmitter and radio play~\cite{yakura2018robust}. This classification helps in understanding the pathways through which adversarial audio can impact VCS, indicating the need for varied defensive strategies.

\subsubsection{Basic Idea of Designing Adversarial Samples}
Adversarial audio samples are essentially a fusion of benign audio with carefully crafted adversarial perturbations. The primary challenge in the industry is to ensure that these perturbations, when added to benign audio, remain imperceptible to human users while effectively leading the VCS to produce erroneous outputs~\cite{li2019adversarial}. This task can be conceptualized as an optimization problem where the attacker's goal is to minimally alter the audio to cause misclassification by the VCS.

The process of designing such attacks in the audio domain, similar to other domains, often involves training neural networks~\cite{chen2023asurvey}. However, the methods used in the image recognition domain, the most prevalent area of adversarial attacks application, are not directly transferable to audio due to fundamental differences. In audio, perturbations are added in the temporal domain requiring higher precision compared to the spatial domain in images~\cite{alzantot2018did, carlini2018audio, vadillo2019universal, abdullah2021hear}.

The optimization problem in this context is expressed as follows:
\begin{equation}
    \label{optimization}
    F(x,\delta, y) = \min_{\|\delta\|}l_m(f(x + \delta), y) + c \cdot\|\delta\|
\end{equation}
Here, $x$ represents the benign audio sample, $\delta$ the added perturbation, and $y$ the label output by the model $f(\cdot)$. The loss function $l_m(\cdot)$ measures the discrepancy between the model's actual output $f(x+\delta)$ and the label $y$, post-perturbation. The term $c \cdot\|\delta\|$ quantifies the perturbation's impact on the original audio, where $c$ is a balancing coefficient between attack effectiveness and audio quality. Higher values of $c$ indicate substantial degradation in audio quality, potentially compromising the original content's integrity.

In untargeted attacks, the attacker seeks the smallest $\delta$ that drastically alters the output $f(x+\delta)$ from $y$ (given $f(x)=y$). Conversely, in targeted attacks, the goal is to minimize $\delta$ such that $f(x+\delta)$ closely matches a specified target output $y'$ (where $f(x)\neq y$).

To tackle the optimization problem inherent in designing adversarial attacks in the preprocessing layer of VCS, several methods, primarily centered around gradient descent and deep learning techniques, have been proposed in the industry.

\emph{Gradient Descent (GD).}
Gradient Descent (GD) is a primary method for creating adversarial audio, widely studied in current research~\cite{cisse2017houdini, Iter2017GeneratingAE, yakura2018robust}. GD optimizes adversarial samples by iteratively calculating the local minimum of a loss function using partial derivatives and the chain rule. Typically utilized in white-box attack scenarios, GD requires an understanding of the target loss function's specifics. While increasing the number of iterations can enhance the success rate of the attack, it may also compromise the adversarial audio's quality~\cite{schonherr2018adversarial}.

\emph{Deep Learning-based Methods.}
To overcome the time and computational resource constraints of iterative GD, deep learning models, such as Recurrent Neural Networks (RNNs)~\cite{chang2020audio}, are employed. These pre-trained models simulate the gradient descent process, enabling real-time generation of adversarial samples, thus enhancing efficiency.

\emph{Fast Gradient Sign Method (FGSM).}
FGSM, introduced by Goodfellow et al.~\cite{goodfellow2014explaining}, offers a faster approach to generate adversarial samples. It involves adding perturbations in the gradient direction of the original audio using a sign function, requiring only a single gradient calculation and addition/subtraction operation. Proven effective in the audio domain, FGSM can efficiently produce adversarial samples~\cite{kreuk2018fooling, li2020adversarial}.

\emph{Transferability of Adversarial Samples.}
Given the limited access to commercial VCS models, attackers often rely on gray-box or black-box approaches, relying on querying the model and interpreting outputs. Transferability of adversarial samples between models becomes crucial in such scenarios. Studies in computer vision have shown the effectiveness of transferring adversarial samples across different models~\cite{fawzi2018adversarial, szegedy2013intriguing}. In audio, the success of transfer-based attacks is linked to the similarity between the parameters and structures of different models~\cite{chen2020devil}. Xu et al.~\cite{li2020adversarial} demonstrated this by successfully executing a transfer-based adversarial attack between i-vector and x-vector models, exploiting their structural similarities.

\subsubsection{The Characteristics of Adversarial Attacks in Preprocessing Layer}
A key characteristic of adversarial attacks in the preprocessing layer is the indirect addition of perturbations. Rather than directly altering the audio signal, attackers inject perturbations into neural network features, such as those obtained from Short-Time Fourier Transform (STFT) or Mel Frequency Cepstral Coefficients (MFCC). For STFT-based approaches, benign audio is first converted into a spectrogram, and adversarial perturbations are added at this stage, followed by inverse transformation to generate adversarial audio~\cite{chang2020audio, cisse2017houdini}. In contrast, perturbing MFCC features, while mitigating some challenges, results in a significant quality reduction in the adversarial audio due to the lossy nature of inverse MFCC transformation~\cite{du2020sirenattack, Iter2017GeneratingAE, yakura2018robust}.

\subsubsection{Defense Schemes against Adversarial Attacks.}
Adversarial training, a method proven effective in image processing, involves integrating adversarial examples into training data to enhance model robustness~\cite{madry2017towards}. However, this approach has limitations: 
\begin{enumerate}
    \item It can lead to decreased accuracy in predicting legitimate examples due to label leakage~\cite{wang2021adversarial}.
    \item The effectiveness of adversarial training depends on the availability of a diverse set of known adversarial examples, making it less effective against novel attacks~\cite{gong2018overview}.
    \item This method increases training iterations, thereby raising the model's training cost.
\end{enumerate}

Given that these attacks often require in-depth knowledge of the VCS's preprocessing techniques, they are generally categorized as white-box attacks. A counter-strategy involves modifying the targeted data preprocessing methods. For example, if attacks primarily target MFCC processing, alternative feature extraction techniques like neural networks or other methods may be employed~\cite{abdel2014convolutional, hai2003improved}. However, introducing new neural network modules to the preprocessing layer may inadvertently create new vulnerabilities within the system.

\begin{table*}[htbp]
    \scriptsize
    \renewcommand\arraystretch{1.9}
    \centering
    \caption{Comparison of voice-synthesis attacks and adversarial attacks in preprocessing layer}
    \label{tab:preprocessing layer}
    \begin{threeparttable}
      \begin{tabular}{c|c|c|c|c|c|c|c|c|c}
      \hline
      \multirow{2}{*}{\makecell[c]{Attack \\ Type}} & \multirow{2}{*}{Paper} & \multirow{2}{*}{\makecell[c]{Attack \\ Method}} & \multirow{2}{*}{Goal} & \multicolumn{1}{c|}{\multirow{2}{*}{\makecell[c] {Attack Target \\ (Category)}}} & \multicolumn{1}{c|}{\multirow{2}{*}{\makecell[c]{Model \\ Knowledge}}} & \multirow{2}{*}{\makecell[c]{\makecell[c]{Attack \\ Vector}}} & \multicolumn{1}{c|}{\multirow{2}{*}{\makecell[c]{Cost \\ (Time)}}} & \multicolumn{2}{c}{Performance} \\
  \cline{9-10}          &       &       &       &       &       & &       & Success & Distance \\
      \hline
      \multicolumn{1}{c|}{\multirow{7}{*}{\begin{sideways}\makecell[c]{Voice-synthesis \\ Attacks}\end{sideways}}} & \makecell[c]{\cite{bispham2019nonsense}} & RCP\tnote{1}    & Targeted & Google Assist (ASR)  & Black & WAA    & n.a     & n.a     & n.a \\
  \cline{2-10}          & \multirow{3}{*}{\makecell[c]{\cite{abdullah2019practical}}} & \multirow{3}{*}{SP\tnote{2}}   & \multirow{3}{*}{Targeted} & \multirow{2}{*}{10 models (ASR)} & \multirow{3}{*}{Black} & WTA & n.a     & 100\%  & n.a \\
  \cline{7-10}          &                       &       &          &              &    & WAA  & n.a     & 80\%   & 1 foot \\
  \cline{5-5}\cline{7-10}          &                       &       &          & 2 models (SV)              &    & WTA  & n.a     & 100\%   & n.a \\
  \cline{2-10}          & \multirow{2}{*}{\makecell[c]{\cite{carlini2016hidden}}} & GD    & \multirow{2}{*}{Targeted} & CMU Sphinx (ASR) & White & \multirow{2}{*}{WAA} & 32h   & 82\%  & 0.5m \\
  \cline{3-3}\cline{5-6}\cline{8-10}          &       & IMFCC\tnote{3} &       & Google Now (ASR) & Black &       & n.a     & 80\%  &  $\leq 3.5$m \\
  \cline{2-10}          & \makecell[c]{\makecell[c]{\cite{vaidya2015cocaine}}} & IMFCC & Targeted & Google Now (ASR) & Black & WAA    & n.a     & n.a     & n.a \\
      \hline
      \multicolumn{1}{c|}{\multirow{15}{*}{\begin{sideways}Adversarial Attacks\end{sideways}}} & \multirow{2}{*}{\makecell[c]{\cite{cisse2017houdini}}} & \multirow{2}{*}{GD} & \multirow{2}{*}{Untargeted} & DeepSpeech-2 (ASR) & White & \multirow{2}{*}{WTA} & \multirow{2}{*}{n.a} & \multirow{2}{*}{n.a} & \multirow{2}{*}{n.a} \\
  \cline{5-6}          &       &       &       & Google Voice (ASR) & Black &       &       &       &  \\
  \cline{2-10}          & \makecell[c]{\cite{Iter2017GeneratingAE}} & GD    & Targeted & WaveNet (ASR) & White & WTA    & n.a     & n.a     & n.a \\
  \cline{2-10}          & \multirow{2}{*}{\makecell[c]{\cite{yakura2018robust}}} & \multirow{2}{*}{GD}    & \multirow{2}{*}{Targeted} & \multirow{2}{*}{DeepSpeech (ASR)} & \multirow{2}{*}{White} & WWA & \multirow{2}{*}{n.a}     & 50$\sim $100\% & \multirow{2}{*}{$\leq 0.5$m} \\
  \cline{7-7}\cline{9-9}          &                       &       &          &              &    & WAA  &      & 60$\sim $90\%   &   \\
  \cline{2-10}          & \makecell[c]{\makecell[c]{\cite{chang2020audio}}} & RNN   & Targeted & KWS (ASR) & White & WAA    & $\thickapprox 0.096$s & 84.30\% & $\leq 4$m \\
  \cline{2-10}          & \multirow{2}{*}{\makecell[c]{\makecell[c]{\cite{kreuk2018fooling}}}} & \multirow{2}{*}{FGSM}   & \multirow{2}{*}{Untargeted} & \multirow{2}{*}{DNN End-to-end (SV)} & White& \multirow{2}{*}{WTA}  & \multirow{2}{*}{n.a} & 64$\sim $94\%(FAR)\tnote{4} & \multirow{2}{*}{n.a} \\
  \cline{6-6}\cline{9-9} &                    &       &          &  & Black &     &    & 46\%(FAR) &\\
  \cline{2-10}          & \multirow{3}{*}{\makecell[c]{\makecell[c]{\cite{li2020adversarial}}}} & \multirow{2}{*}{FGSM}   & \multirow{3}{*}{Untargeted} & \multirow{2}{*}{GMM i-vector (SV)} & White & \multirow{3}{*}{WTA}  & \multirow{3}{*}{n.a} & 99.99\%(FAR) & \multirow{3}{*}{n.a} \\
  \cline{6-6}\cline{9-9} &                                                                      &                        &                             &                                    & Black &                       &                    & \textless 70\%(FAR) &                       \\
  \cline{3-3}\cline{5-6}\cline{9-9} &                    & Transfer      &          & DNN x-vector (SV) & Black  &     &    & \textless 60\%(FAR) &\\
  \cline{2-10}          & \makecell[c]{\cite{marras2019adversarial}} & GD    & Targeted & VGGVox (SV) & White & WTA    & n.a     & \textless 80\%     & n.a \\
  \cline{2-10}          & \makecell[c]{\cite{wang2020inaudible}} & GD    & Targeted & DNN x-vector (SV) & White & WTA    & n.a     & \textless 98.5\%     & n.a \\
  \cline{2-10}         & \multirow{2}{*}{\makecell[c]{\cite{guo2022specpatch}}} & \multirow{2}{*}{\makecell[c]{RCP}} & \multirow{2}{*}{Targeted} & \multirow{2}{*}{\makecell[c]{DeepSpeech2 (ASR)}} & \multirow{2}{*}{\makecell[c]{Black}}  & WTA & \multirow{2}{*}{\makecell[c]{n.a}} & 100\% & n.a \\
  \cline{7-7}\cline{9-10}         &    &    &     &    &    & WAA   &    & 80\%   & 1m \\

      \hline
      \end{tabular}
      \begin{tablenotes}
        \item ${}^1$ RCP: replacing the consonant phonemes. ${}^2$ SP: signal processing. ${}^3$ IMFCC: inverse MFCC. ${}^4$ FAR: false acceptance rate.
      \end{tablenotes}
    \end{threeparttable}
  \end{table*}%
  
\section{Kernel Layer}
In VCS, the kernel layer plays a crucial role in decoding audio feature vectors processed in the preprocessing layer. This decoding is traditionally achieved through an acoustic model, which translates feature vectors into phonemes, and a language model, predicting word sequence probabilities to output the most probable sentence. Historically, decoding methods have included Hidden Markov Models (HMM), Recurrent Neural Networks (RNN), and N-gram models. However, the rise of deep learning technologies has led to advanced models like Wav2Letter (CNNs-based), Kaldi (DNN-HMM), and CMU Sphinx (GMM-HMM) becoming mainstream due to their enhanced accuracy in speech recognition~\cite{nassif2019speech}. Despite these advancements, the kernel layer remains vulnerable to adversarial attacks that manipulate the decoding process and spoofing attacks, where attackers mimic legitimate audio inputs to deceive the system, highlighting the need for robust security measures in this critical component of VCS.

\subsection{Spoofing Attacks in Kernel Layer}
Spoofing attacks in the kernel layer of VCS primarily target the Speaker Verification (SV) component. These attacks are executed with the intention of impersonating a target speaker to gain unauthorized access or privileges within the VCS system~\cite{ergunay2015vulnerability, evans2013spoofing}. Commonly, spoofing attacks are categorized into four types: replay, impersonation, speech synthesis, and voice conversion.

\subsubsection{Replay}
A replay attack involves recording a victim's voice using a recording device and then replaying this captured audio to deceive the SV system. This method is straightforward yet highly effective, posing a significant threat to various SV systems due to its simplicity and reliability. The effectiveness of replay attacks has been demonstrated against many types of SV systems~\cite{lindberg1999vulnerability, villalba2010speaker, villalba2011detecting, yoon2020new}.

\subsubsection{Impersonation}
Impersonation attacks are executed by an attacker manually mimicking the voice patterns and speech behavior of the target individual, without the aid of any computerized devices. These attacks aim to trick the SV component into falsely recognizing the attacker as the target. Research has indicated that even non-professional impersonators can be successful, particularly if they are familiar with the target's voice pattern or possess a naturally similar voice~\cite{lau2005testing, lau2004vulnerability}. Professional impersonators often try to replicate rhythmic features of the target's speech to enhance the deception~\cite{farrus2008vulnerable}. However, it has been observed that significant differences between the impersonator's and the target's natural voice can still be detected by SV systems~\cite{hautamaki2013vectors, mariethoz2005can}.
\subsubsection{Speech Synthesis}
Speech Synthesis (SS) is a technology that aims to generate speech resembling the victim's voice from input text, distinct from standard text-to-speech (TTS) applications. The goal is to deceive the SV component by impersonating the victim. Traditional SS techniques include unit selection~\cite{hunt1996unit}, statistical parametric modeling~\cite{zen2009statistical}, and hybrid approaches combining both~\cite{qian2012unified}. Advancements in deep learning have led to the use of neural networks in SS, such as Generative Adversarial Networks (GAN) for enhancing speech quality~\cite{binkowski2019high, pascual2017segan, saito2017statistical}, and specific models like WaveNet~\cite{oord2018parallel, vanwavenet} and Tacotron~\cite{wang2017tacotron} for victim-specific speech generation.

\subsubsection{Voice Conversion}
Voice Conversion (VC) differs from SS as it transforms the attacker's speech to match the victim's speech patterns. Traditional VC methods used statistical models like Hidden Markov Models (HMM)~\cite{kim1997hidden}, Gaussian Mixture Models (GMM)~\cite{stylianou1998continuous}, Principal Component Analysis (PCA)~\cite{wilde2004probabilistic}, and Non-negative Matrix Factorization (NMF)~\cite{zhang2015non}. Modern approaches leverage neural networks, including Artificial Neural Networks (ANN)~\cite{desai2009voice}, WaveNet, and GANs. Recent research has focused on challenges such as low-resource languages~\cite{GHORBANDOOST2015113, jannati2018part} and cross-lingual voice conversion~\cite{wahlster2013verbmobil, zhou2019cross}. Innovative techniques involving specially designed tubes have been explored to bypass liveness detection systems by modifying the attacker's voice to resemble the victim's~\cite{ahmed2023tubes}.

\subsubsection{Twins}
Identical twins present a unique challenge in SV due to their highly similar speech characteristics. While SV systems are generally effective in distinguishing between different individuals, they struggle to differentiate between identical twins. Studies have indicated that twins share similar speech signal patterns, pitch contours, formant contours, and spectrograms~\cite{patil2009variable, kersta1970spectrographic}, leading to higher False Acceptance Rates (FAR) in SV. Incidents have been reported where twins were able to access each other's accounts by bypassing SV checks~\cite{noauthor_2017-vv}.
\subsubsection{Defense Schemes against Spoofing Attacks.}
Defense strategies against spoofing attacks in VCS primarily focus on identifying distortions in audio signals during transcription. 

\textbf{For Replay Attacks:} Techniques have been developed to distinguish between original and replayed audio by analyzing far-field recordings characteristic of distant attacks~\cite{villalba2011detecting, villalba2011preventing}. Another approach involves maintaining a database of previous audio recordings in VCS and calculating similarity scores for new inputs to detect replays~\cite{shang2010score}.

\textbf{For Speech Synthesis (SS):} Traditional SS detection methods often rely on understanding acoustic differences based on specific SS algorithms, such as analyzing spectral parameter ranges~\cite{satoh2001robust} and variations in mel-cepstral coefficients~\cite{chen2010speaker}. Differences in pitch patterns and vocal tract models between human and synthesized audio offer additional detection markers~\cite{ogihara2005discrimination, de2012synthetic, de2012evaluation, wu2012detecting, wu2012study}.

\textbf{For Voice Conversion (VC):} Defenses against VC attacks draw inspiration from SS defense strategies, with studies showing that SVM classifiers based on super-vectors demonstrate inherent robustness against VC~\cite{alegre2012spoofing, alegre2013spoofing}. Analyzing pairwise distances between continuous feature vectors in human versus VC-generated audio has also proven effective.

\textbf{Machine Learning-based Approaches:} State-of-the-art methods involve using machine learning models to discern spectral differences between human and generated audio~\cite{gong2020detecting, jelil2018exploration, kamble2018effectiveness, patil2019energy, wang2017feature, witkowski2017audio}. The ASVspoof challenges have been instrumental in driving advancements in anti-spoofing technology, with RawNet2 emerging as a successful model in this domain~\cite{tak2021end, wu2015asvspoof, kinnunen2017asvspoof, todisco2019asvspoof, yamagishi2021asvspoof}. Additionally, research combining articulatory phonetics to reconstruct vocal tracts from audio signals has shown promise in distinguishing between human and machine sources~\cite{blue2022you}. A novel approach proposed in \cite{deng2023catch} shifts focus from preemptively rejecting VC-generated audio to reconstructing the source speaker's voiceprint for attacker identification.

\subsection{Adversarial Attacks in Kernel Layer}
The kernel layer of VCS, much like the preprocessing layer, is vulnerable to adversarial attacks. However, the attack methodology in the kernel layer typically involves directly adding adversarial perturbations to the audio waveforms of legitimate audio samples. This direct manipulation of audio waveforms distinguishes these attacks from those in the preprocessing layer. Adversarial attacks in the kernel layer pose a significant challenge to audio classifiers, often leading to misclassifications and subsequent security breaches. A detailed summary and comparative analysis of recent advancements in adversarial attacks targeting the kernel layer can be found in Table \ref{tab:kernel layer}. This table provides insights into the various techniques and strategies employed in these attacks, underlining their potential impact on VCS performance and security.

\subsubsection{Adversarial Examples Crafting}
In the kernel layer of VCS, the shift towards black-box models has made traditional gradient-based adversarial sample generation methods like Gradient Descent (GD) and Fast Gradient Sign Method (FGSM) less effective. To address this, researchers have developed gradient-agnostic algorithms for crafting adversarial examples.

\emph{Genetic Algorithm (GA).} GA employs a search-based approach, iterating through potential samples to deceive machine learning models. By using techniques like crossover and mutation, GA refines search results, prioritizing samples with higher fitness scores for subsequent iterations. This method is suitable for gray-box or black-box attacks due to its reliance solely on model output rather than gradient information~\cite{taori2019targeted, han2019nickel, khare2019adversarial, wu2019semi, wang2020towards}. However, GA's iterative nature leads to high computational resource demands.

\emph{Particle Swarm Optimization (PSO).} Originating from swarm intelligence, PSO mimics collective animal behaviors to search for optimal solutions~\cite{kennedy1995particle}. In adversarial example generation, samples are treated as swarm particles, moving towards the most effective position in the search space iteratively~\cite{du2020sirenattack}. PSO benefits from fast convergence and does not require gradient information, making it efficient for gray-box and black-box attacks.

\emph{Ensemble Algorithm.} Chen et al.~\cite{chen2021real} introduced a novel black-box approach combining the Natural Evolution Strategy (NES)~\cite{wierstra2014natural} and the Basic Iterative Method (BIM)~\cite{kurakin2018adversarial}. NES estimates the gradient of the black-box model, and this estimated gradient is then used in BIM to generate adversarial examples iteratively. This method efficiently crafts adversarial examples suitable for black-box settings.

\emph{Deep Learning-based Methods.} Approaches like the Gated Convolutional Auto-encoder (GCA)~\cite{shamsabadi2021foolhd}, Adversarial Transformation Network (ATN)~\cite{li2021learning}, and Generative Network (GN)~\cite{li2020universal} employ deep learning for adversarial example generation. GN is effective in gray-box model attacks, while ATN significantly reduces the time required to generate adversarial examples.

\subsubsection{Defense Schemes against Adversarial Attacks.}
While the defense methods used in the preprocessing layer of VCS are also applicable to the kernel layer, the nature of attacks targeting the kernel layer demands additional strategies. Since these attacks are typically directed at the model itself, the transferability of adversarial examples between different models tends to be limited. Zeng et al.~\cite{zeng2019multiversion} observed that adversarial examples crafted for one specific model often yield different prediction results when applied to another model, a discrepancy not seen with normal audio signals. This unique behavior of adversarial examples provides a basis for detecting and countering attacks in the kernel layer. By exploiting the differences in model responses to adversarial versus normal audio, defense mechanisms can be designed to effectively identify and mitigate adversarial attacks targeting this crucial layer of VCS.

\begin{table*}[htbp]
  \scriptsize
    \renewcommand\arraystretch{1.35}
    \centering
    \caption{Comparison of voice-synthesis attacks and adversarial attacks in kernel layer}
    \begin{threeparttable}
      \begin{tabular}{c|c|c|c|c|c|c|c|c|c}
      \hline
      \multicolumn{1}{c|}{\multirow{2}{*}{\makecell[c]{Attack \\ Type}}} & \multirow{2}{*}{Paper} & \multirow{2}{*}{\makecell[c]{Attack \\ Method}} & \multirow{2}{*}{Goal} & \multicolumn{1}{c|}{\multirow{2}{*}{\makecell[c]{Attack Target \\ (Category)}}} & \multicolumn{1}{c|}{\multirow{2}{*}{\makecell[c]{Model \\ Knowledge}}} & \multicolumn{1}{c|}{\multirow{2}{*}{\makecell[c]{Attack \\ Vector}}} & \multicolumn{1}{c|}{\multirow{2}{*}{\makecell[c]{Cost \\ (Time)}}} & \multicolumn{2}{c}{Performance} \\
  \cline{9-10}          &       &       &       &       &       &       &       & Success & Distance \\
      \hline
      \multicolumn{1}{c|}{\multirow{60}{*}{\begin{sideways}Adversarial Attacks\end{sideways}}} & \multirow{3}{*}{\makecell[c]{~\cite{yuan2018commandersong}}} & \multirow{3}{*}{GD} & \multirow{3}{*}{Targeted} & \multirow{2}{*}{Kaldi (ASR)} & \multirow{2}{*}{White} & WTA   & \textless  1.9h & 100\% & n.a \\
  \cline{7-10}          &       &       &       &       &       & WAA   & \textless  2h  & 96\%  & 1.5m \\
  \cline{5-10}          &       &       &       & iFLYTEK (ASR) & Black & WAA   & n.a & 66\%  & n.a \\
  \cline{2-10}          & \makecell[c]{~\cite{schonherr2019adversarial}} & GD    & Targeted & Kaldi (ASR) & White & WTA   & 2min  & 98\%  & n.a \\
  \cline{2-10}          & \multirow{2}{*}{\makecell[c]{~\cite{qin2019imperceptible}}} & \multirow{2}{*}{GD} & \multirow{2}{*}{Targeted} & \multirow{2}{*}{Lingvo (ASR)} & \multirow{2}{*}{White} & WTA   & \multirow{2}{*}{n.a} & 100\% & \multirow{2}{*}{n.a} \\
  \cline{7-7}\cline{9-9}          &       &       &       &       &       & WAA   &       & 60\%  &  \\
  \cline{2-10}          & \makecell[c]{~\cite{taori2019targeted}} & GD+GA & Targeted & DeepSpeech (ASR) & Black & WTA   & n.a     & 35\%  & n.a \\
  \cline{2-10}          & \makecell[c]{~\cite{neekhara2019universal}} & GD    & Targeted & 2 model (ASR) & White & WTA   & n.a     & 89.6\% & n.a \\
  \cline{2-10}          & \makecell[c]{~\cite{kwon2019selective}} & GD    & Targeted & DeepSpeech (ASR) & White & WTA   & 1h    & 91.7\% & n.a \\
  \cline{2-10}          & \makecell[c]{~\cite{li2019adversarial}} & GD    & Targeted & Amazon Alexa (ASR) & Gray  & WAA   & n.a     & 50\%  & 7.7ft \\
  \cline{2-10}          & \makecell[c]{~\cite{gong2019real}} & RL+RNN & Untargeted & KWS (ASR) & Gray  & WTA   & 0.15s & 43.5\% & n.a \\
  \cline{2-10}          & \makecell[c]{~\cite{liu2020weighted}} & GD    & Targeted & DeepSpeech (ASR) & White & WTA   & 4-5min & n.a     & n.a \\
  \cline{2-10}          & \makecell[c]{~\cite{schonherr2020imperio}} & GD    & Targeted & Kaldi (ASR) & White & WAA   & 80min & 25\%  & 3m \\
  \cline{2-10}          & \makecell[c]{~\cite{szurley2019perceptual}} & GD    & Targeted & DeepSpeech (ASR) & White & WAA   & n.a     & n.a     & 6 in \\
  \cline{2-10}          & \makecell[c]{~\cite{han2019nickel}} & GA    & Untargeted & Google Cloud (ASR) & Black & WTA   & n.a     & 86\%  & n.a \\
  \cline{2-10}          & \multirow{2}{*}{\makecell[c]{~\cite{khare2019adversarial}}} & \multirow{2}{*}{GA} & Untargeted & \multirow{2}{*}{2 models (ASR)} & \multirow{2}{*}{Black} & \multirow{2}{*}{WTA} & \multirow{2}{*}{n.a} & \multirow{2}{*}{n.a} & \multirow{2}{*}{n.a} \\
  \cline{4-4}          &       &       & Targeted &       &       &       &       &       &  \\
  \cline{2-10}          & \makecell[c]{~\cite{wu2019semi}} & GA    & Targeted & Kaldi (ASR) & Gray  & WTA   & n.a     & 90\%  & n.a \\
  \cline{2-10}          & \makecell[c]{~\cite{gong2019audidos}} & GD    & Untargeted & 2 models (ASR) & White & WAA   & n.a     & 78\%  & 30cm \\
  \cline{2-10}          & \multirow{2}{*}{\makecell[c]{~\cite{chen2020devil}}} & GD    & \multirow{2}{*}{Targeted} & 4 models (ASR) & \multirow{2}{*}{Black} & \multirow{2}{*}{WAA} & \multirow{2}{*}{n.a} & 98\%  & 2m \\
  \cline{3-3}\cline{5-5}\cline{9-10}          &       & Transfer &       & Apple Siri (ASR) &       &       &       & n.a     & n.a \\
  \cline{2-10}          & \makecell[c]{~\cite{chen2020metamorph}} & GD    & Targeted & DeepSpeech (ASR) & White & WAA   & 5-7h  & 90\%  & 6m \\
  \cline{2-10}          & \makecell[c]{~\cite{wang2020targeted}} & GAN   & Targeted & 2 models (ASR) & White & WTA   & 0.009s & 92.33\% & n.a \\
  \cline{2-10}          & \makecell[c]{~\cite{wang2020towards}} & GA    & Targeted & DeepSpeech (ASR) & Gray  & WTA   & n.a     & 98\%  & n.a \\
  \cline{2-10}          & \multirow{2}{*}{\makecell[c]{~\cite{li2020advpulse}}} & \multirow{2}{*}{GD} & \multirow{2}{*}{Targeted} & KWS (ASR) & \multirow{2}{*}{White} & \multirow{2}{*}{WAA} & \multirow{2}{*}{n.a} & 89.9\% & \multirow{2}{*}{3m} \\
  \cline{5-5}\cline{9-9}          &       &       &       & DNN x-vector (SV) &       &       &       & 89.3\% &  \\
  \cline{2-10}          & \multirow{2}{*}{\makecell[c]{~\cite{abdullah2021hear}}} & \multirow{2}{*}{SP} & \multirow{2}{*}{Untargeted} & 7models (ASR) & \multirow{2}{*}{Black} & \multirow{2}{*}{WTA} & \multirow{2}{*}{n.a} & 100\% & \multirow{2}{*}{n.a} \\
  \cline{5-5}\cline{9-9}          &       &       &       & Microsoft Azure (SV) &       &       &       & n.a     &  \\
  \cline{2-10}          & \multirow{3}{*}{\makecell[c]{~\cite{du2020sirenattack}}} & \multirow{3}{*}{PSO}   & Untargeted & DeepSpeech (ASR) & White & \multirow{3}{*}{WTA} & \multirow{3}{*}{n.a} & 100\% & 1201.7s \\
  \cline{4-6}\cline{9-10}          &       &       & \multirow{2}{*}{Targeted} & 7models (ASR) & \multirow{2}{*}{Black} &       &       & 95.25\% & 100.7s \\
  \cline{5-5}\cline{9-10}          &       &       &       & 7models (SV) &       &       &       & 99.45\% & 376.4s \\
  \cline{2-10}          & \multirow{5}{*}{\makecell[c]{~\cite{chen2021real}}} & \multirow{3}{*}{NES+BIM} & Untargeted & \multirow{2}{*}{3 models (SV)} & \multirow{5}{*}{Black} & \multirow{2}{*}{WAA} & \multirow{5}{*}{n.a} & 100\% & \multirow{2}{*}{2m} \\
  \cline{4-4}\cline{9-9}          &       &       & Targeted &       &       &       &       & 95\%  &  \\
  \cline{4-5}\cline{7-7}\cline{9-10}          &       &       & Targeted & Talentedsoft (SV) &       & WTA   &       & 100\% & n.a \\
  \cline{3-5}\cline{7-7}\cline{9-10}          &       & \multirow{2}{*}{Transfer} & Untargeted & \multirow{2}{*}{Microsoft Azure (SV)} &       & \multirow{2}{*}{WAA} &       & 57\%  & \multirow{2}{*}{2m} \\
  \cline{4-4}\cline{9-9}          &       &       & Targeted &       &       &       &       & 34\%  &  \\
  \cline{2-10}          & \makecell[c]{~\cite{wang2019adversarial}} & FGSM  & Untargeted & GE2E (SV) & White & WTA   & n.a     & n.a     & n.a \\
  \cline{2-10}          & \makecell[c]{~\cite{xie2020real}} & SP   & Targeted & DNN x-vector (SV) & White & WAA   & 0.015s & 90\%  & 5m \\
  \cline{2-10}          & \multirow{4}{*}{\makecell[c]{~\cite{li2020practical}}} & \multirow{4}{*}{FGSM} & \multirow{2}{*}{Untargeted} & \multirow{4}{*}{DNN x-vector (SV)} & \multirow{4}{*}{White} & WTA   & \multirow{4}{*}{n.a} & 99\%  & n.a \\
  \cline{7-7}\cline{9-10}          &       &       &       &       &       & WAA   &       & 100\% & 1m \\
  \cline{4-4}\cline{7-7}\cline{9-10}          &       &       & \multirow{2}{*}{Targeted} &       &       & WTA   &       & 96.01\% & n.a \\
  \cline{7-7}\cline{9-10}          &       &       &       &       &       & WAA   &       & 50\%  & 1m \\
  \cline{2-10}          & \multirow{2}{*}{\makecell[c]{~\cite{li2020universal}}} & \multirow{2}{*}{GN} & Untargeted & \multirow{2}{*}{SineNet (SV)} & \multirow{2}{*}{Gray} & \multirow{2}{*}{WTA} & \multirow{2}{*}{n.a} & \multirow{2}{*}{n.a} & \multirow{2}{*}{n.a} \\
  \cline{4-4}          &       &       & Targeted &       &       &       &       &       &  \\
  \cline{2-10}          & \multirow{3}{*}{\makecell[c]{~\cite{zhang2020voiceprint}}} & \multirow{2}{*}{GD} & \multirow{3}{*}{Targeted} & VGGVox (SV) & Gray  & \multirow{2}{*}{WTA} & \multirow{3}{*}{n.a} & 100\% & \multirow{3}{*}{n.a} \\
  \cline{5-6}\cline{9-9}          &       &       &       & Microsoft Azure (SV) & \multirow{2}{*}{Black} &       &       & 75\%  &  \\
  \cline{3-3}\cline{5-5}\cline{7-7}\cline{9-9}          &       & Transfer &       & Apple Homekit (SV) &       & WAA   &       & 67.5\% &  \\
  \cline{2-10}          & \makecell[c]{~\cite{li2021learning}} & ATN   & Targeted & SineNet (SV) & White & WTA   & 0.042RTF\tnote{1} & n.a     & n.a \\
  \cline{2-10}          & \multirow{2}{*}{\makecell[c]{~\cite{shamsabadi2021foolhd}}} & \multirow{2}{*}{GCA} & Untargeted & \multirow{2}{*}{DNN x-vector (SV)} & \multirow{2}{*}{White} & \multirow{2}{*}{WTA} & \multirow{2}{*}{n.a} & 99.6\% & \multirow{2}{*}{n.a} \\
  \cline{4-4}\cline{9-9}          &       &       & Targeted &       &       &       &       & 99.2\% &  \\
  \cline{2-10}         & \multirow{4}{*}{\makecell[c]{~\cite{yu2023smack}}} & \multirow{4}{*}{GA + GE} &\multirow{4}{*}{Targeted} &\multirow{2}{*}{5 models (ASR)} &\multirow{4}{*}{Black}  & WTA & 474s & 87.7\% & n.a \\
  \cline{7-10}         &    &    &    &    &    & WAA   & n.a   & 43.3\%   & 200cm \\
  \cline{5-5}\cline{7-10}   &    &    &    &\multirow{2}{*}{2 models (SV)}    &    & WTA   & 525s   & 99.2\%   & n.a  \\
  \cline{7-10}         &    &    &    &    &    & WAA   & n.a   & 35.8\%   & 200cm \\
  \cline{2-10}         & \multirow{3}{*}{~\cite{zheng2021black}} & \multirow{2}{*}{CC-CMA-ES} & \multirow{3}{*}{Targeted} & 5 models (ASR) & \multirow{3}{*}{Black}  & \multirow{2}{*}{WTA} & \multirow{3}{*}{n.a} & \multirow{2}{*}{100\%} & \multirow{2}{*}{n.a} \\
  \cline{5-5} &                                      &                            &                           & 2 models (SV) &                          &                      &            &    &         \\
  \cline{3-3}\cline{5-5}\cline{7-7}\cline{9-10}         &                                  & DNN                    &                           & 5 models (ASR) &              & WAA                  &     & 60\% & 15cm \\
  \cline{2-10}         & ~\cite{liu2022evil} & GE & Targeted & 4 models (ASR) & Black  & WNA & 362.18s & 87.7\% & n.a \\
    \hline
      \end{tabular}%
      \begin{tablenotes}
        \item[1] RTF: the ratio of the processing time to the input duration
      \end{tablenotes}
    \end{threeparttable}
    \label{tab:kernel layer}%
  \end{table*}%

  \begin{tcolorbox}
    Remark 1: Communication and video conferencing through IP have become the main way of community communication nowadays. In \cite{liu2022evil}, researchers found that due to the special characteristics of IP channels, 
    audio signal transmission can cause signal attenuation, random channel noise, and other interference, 
    which can affect the effectiveness of audio adversarial samples. 
    Therefore, in future research, it is necessary to focus on investigating the potential impact of malicious audio in the transmission process over IP channels. 
    In this paper, we refer to attacks that are transmitted through IP channels as Wav-Network-API (WNA).
  \end{tcolorbox}
\section{Service Layer}
The evolution of VCS has led to the integration of third-party services, enhancing functionality beyond basic speech content recognition or identity verification. Examples include Amazon Echo and Google Assistant. While this expansion addresses user demands, it also introduces heightened security risks. This section explores the potential vulnerabilities and threats posed by these third-party integrations in VCS.

\subsection{Squatting Attacks in Service Layer}
Despite mandatory reviews for third-party services in VCS to ensure compliance with security and privacy policies, many services still breach these regulations~\cite{guo2020skillexplorer, su2020you}, and current review processes struggle to identify such violations~\cite{cheng2020dangerous}. This oversight allows attackers to introduce malicious services, posing significant privacy risks. Squatting attacks occur when attackers create services with names resembling legitimate ones, deceiving users into accessing these malicious options instead. There are two primary methods of squatting attacks:

\begin{enumerate}
    \item \textbf{Similar Sounding Service Names:} Attackers may design service names phonetically similar to legitimate ones, leading to accidental activation of the malicious service. For instance, a service named ``Capital Won" could be mistaken for ``Capital One", especially in noisy settings or based on user dialect or gender~\cite{zhang2019dangerous, kumar2018skill}.
    
    \item \textbf{Exploiting Longest Matching Rule:} VCS typically match service names using the longest matching rule. Attackers leverage this and user habits to construct squatting attacks. Since many users add polite terms like ``please" to service commands, a malicious service named ``Capital One Please" could be preferentially activated over the intended ``Capital One" service, as observed in user habit surveys~\cite{zhang2019dangerous}.
\end{enumerate}

These squatting tactics demonstrate the nuanced vulnerabilities introduced by third-party service integration in VCS, highlighting the need for more robust security measures.

\emph{Misunderstanding Attack.}
A misunderstanding attack is a nuanced form of squatting attack where the goal is not to activate a malicious service, but to induce errors within a legitimate service. These attacks exploit the limitations of VCS in accurately interpreting words outside their programmed understanding~\cite{stolk2016conceptual}. For example, Siri might misinterpret ``bank" in the context of ``river bank". Bispham et al. ~\cite{bispham2019nonsense} demonstrated that by altering a single word in a command, VCS can be misled. In a banking service, changing ``money" to ``dough" in the query ``how much money do I have?" could result in unintended information disclosure due to ``dough" being slang for money. The Natural Language Understanding (NLU)'s Intent Classifier, which uses fuzzy matching to accommodate user speech variations, is often the exploitable component in such scenarios~\cite{zhang2019life}.

\subsubsection{Defense Schemes against Squatting Attacks.}
Squatting attacks often involve similar-sounding service names or variants to trick VCS. Zhang et al. ~\cite{zhang2019dangerous} developed a phoneme-based service name scanner to identify suspicious third-party services. Additionally, it is suggested that service platform providers analyze new services for potential naming conflicts or utilize interactive tools for policy violation detection~\cite{kumar2018skill, young2022skilldetective}.

\subsection{Faking Termination in Service Layer}
Users' misunderstandings about service termination protocols in VCS like Alexa and Google Assistant can be exploited. Users often assume that services end automatically after silence or a farewell response. However, malicious services can simulate termination sounds while remaining active, posing significant security risks if users then disclose private information or issue commands. Such deceptive practices highlight the need for greater awareness and more explicit termination protocols in VCS~\cite{zhang2019dangerous}.

\subsubsection{Defense Schemes against Faking Termination.}
To mitigate the risks associated with faking termination in VCS, various studies recommend enhancing service publishing authentication mechanisms~\cite{cheng2020dangerous,guo2020skillexplorer}. Service platform providers should conduct thorough word-based or phoneme-based reviews of new services to avoid confusion with existing ones. Additionally, methods such as voice interaction analysis and backend code review are effective in identifying privacy breaches or discrepancies between a service's functionality and its description. Recent research suggests that context-sensitive detectors could effectively identify instances of faking termination~\cite{zhang2019dangerous}. Notably, a proposed hardening framework for VCS in ~\cite{vaere2023hey} focuses on ensuring that data collected by microphones is used only for local processing unless explicitly permitted by the user, thereby safeguarding against privacy data leakage caused by faking termination.

\section{General Attack Mitigation Strategy} \label{sec:GDS}
In earlier sections, we discussed defense schemes that were primarily focused on countering specific types of attacks in VCS. This section shifts the focus to state-of-the-art general attack mitigation strategies that are capable of defending against more attacks, encompassing different types and layers. These comprehensive strategies are essential for providing a more holistic defense mechanism for VCS systems. We have systematized existing general attack mitigation strategies and evaluated their efficacy against various attack types and characteristics. The comparative analysis of these strategies, including their strengths and limitations in resisting different forms of attacks, is presented in Table \ref{tab:general_compare}. This systematic overview aims to guide researchers and practitioners in selecting and implementing effective, multi-faceted defense strategies for VCS.

\subsection{Liveness Detection}
Liveness detection has emerged as a prevalent defense strategy in VCS, primarily designed to ascertain whether voice commands originate from actual humans. The fundamental premise behind this approach is the recognition that most malicious commands are machine-generated. These commands are typically played through speakers or directly input into the VCS API via audio files, such as WAV files. Unlike these artificially produced commands, genuine human users do not generate voice commands in this manner. Therefore, by identifying the characteristics of human speech, liveness detection aims to filter out these non-human, machine-generated inputs, thereby enhancing the security of VCS.
\emph{Passive Detection.}
In the context of liveness detection for VCS, passive detection plays a crucial role in distinguishing between human-spoken voice commands and those generated by speakers. This is achieved through the analysis of sound characteristics using two main techniques:

\begin{enumerate}
    \item \textbf{Detecting Speaker Characteristics:} Voice commands emitted from speakers often carry unique signal distortions, which are a result of circuit noise inherent to the speaker's hardware. These distortions differ significantly from the patterns found in human speech and can be identified using classifiers trained specifically for this purpose~\cite{ahmed2020void, blue2018hello, gong2018protecting, pradhan2019combating, wang2020differences}. Moreover, the electromagnetic fields generated by speakers during sound emission, owing to the electronic components, can be detected using a magnetometer in smart devices, further aiding in the determination of the voice command's origin~\cite{chen2017you}.

    \item \textbf{Detecting Human Voice Characteristics:} Human speech is produced through a complex physiological process involving the coordinated action of the mouth, vocal tract, vocal cords, and lungs. The airflow from the lungs, passing through the glottis, causes vocal cord vibrations, which are then amplified by resonances in the mouth and vocal tract to form the final sound signal. Identifying features inherent to this process, such as breathing airflow patterns~\cite{mochizuki2018voice, pradhan2019combating, shiota2016voice, wang2019secure, zhou2019hidden}, mouth movements~\cite{meng2018wivo, shang2020voice, zhang2017hearing, yang2023voshield}, and bone vibrations~\cite{feng2017continuous, sahidullah2017robust, shang2018defending, shang2019enabling, shang2020secure, zhang2020viblive}, provides a basis for determining whether a voice command is human-generated. These features can be monitored using microphones, cameras, or other specialized sensors. In practical applications, integrating additional devices or sensors might be necessary to enhance the accuracy and reliability of such verifications.
\end{enumerate}

These passive detection methods are instrumental in bolstering the security of VCS by ensuring that voice commands are genuinely human and not artificially generated or replayed through electronic devices.

\emph{Active Interaction.}
The active interaction defense scheme in VCS involves engaging users in a manner similar to a CAPTCHA to ascertain the authenticity of voice commands. A prevalent form of this scheme is the Challenge-Response mechanism. Upon receiving a voice command, VCS issues a challenge to the user, requiring an appropriate response within a predetermined time frame. Failure to respond correctly within this window leads to the assumption that the command is machine-generated, and thus, the command execution is denied \cite{carlini2016hidden, sugawara2020light, yuan2018all}. While effective in thwarting voice attacks to an extent, this method introduces additional steps for the user, potentially compromising the usability of the VCS.

\begin{tcolorbox}
    \textbf{Remark 2:} Recent research \cite{ahmed2023tubes} has demonstrated that malicious attacks can manipulate a speaker's voice commands into harmful instructions using a specially designed tube. As these instructions are essentially uttered by a real person, they can effectively circumvent traditional liveness detection methods. Future research in VCS security must account for the implications of such attacks on the efficacy of existing liveness detection schemes.
\end{tcolorbox}

\subsection{Audio Conversion}
The conversion of audio in the preprocessing layer of VCS, before it is passed to subsequent layers for further processing, serves as a defensive measure against voice-synthesis and adversarial attacks. This effectiveness stems from the conversion process's ability to disrupt the specific patterns and structures that these attacks aim to exploit or deceive. As a result, the converted audio loses the characteristics intended by the attack, rendering it ineffective. In contrast, benign audio typically exhibits greater resilience to these conversions and is only minimally affected, preserving its integrity while mitigating potential threats.

\emph{Audio Encoding.}
Encoding incoming audio has been shown to effectively diminish the success rate of malicious audio attacks. Utilizing codecs like Advanced Audio Coding (AAC)~\cite{rajaratnam2018speech}, MP3~\cite{andronic2020mp3, das2019adagio, rajaratnam2018speech, zhang2019defending}, Speex~\cite{rajaratnam2018speech}, Opus~\cite{rajaratnam2018speech}, Adaptive Multi-Rate (AMR)~\cite{das2019adagio}, and integrated multiple codecs~\cite{rajaratnam2018isolated, das2019adagio} can provide a substantial level of defense against malicious audio.

\emph{Audio Filtering.}
Voice-synthesis and adversarial attacks rely heavily on precise algorithmic perturbations. Filtering these malicious audio inputs, using methods like median filters~\cite{chen2020devil, du2020sirenattack, yang2018characterizing}, quantization~\cite{yang2018characterizing}, and other noise reduction algorithms~\cite{tamura2019novel, guo2020inor}, effectively destroys these perturbations, thereby safeguarding the VCS from such attacks.

\emph{Audio Downsampling.}
Experiments have demonstrated that downsampling audio to a lower rate and then upsampling it back to a rate suitable for VCS input can effectively mitigate attack impacts. While benign audio remains relatively unaffected, the malicious audio loses its carefully added perturbations, thus failing to achieve the intended effect on the target VCS~\cite{carlini2016hidden, chen2020devil, du2020sirenattack, yang2018characterizing, yuan2018commandersong, tamura2019novel}.

\begin{tcolorbox}
    Remark 3: Audio conversion as a defense strategy exploits the resilience of benign audio to counteract the fragility of malicious audio. This process disrupts the attacker's carefully engineered perturbations. However, in a white-box scenario, attackers aware of the VCS's audio conversion methods can adjust their strategies accordingly, potentially neutralizing the defense's effectiveness.
\end{tcolorbox}

\begin{table*}[htbp]
    \scriptsize
    \renewcommand\arraystretch{2.1}
    \centering
    \caption{Capability  of defending against  different attacks in different layers. 
    (TA: transduction attack, 
    VsA: voice-synthesis attack, 
    AA: adversarial attack, 
    SpA: spoofing attack, 
    SqA: squatting attack, FT: faking termination,  $\bigoplus$ : Defensible, $\bigodot$: Probably Defensible, $\bigotimes$ : Indefensible)}
      \begin{tabular}{c|c|c|c|c|c|c|c|c|c|l}
      \hline
      \multicolumn{2}{c|}{\multirow{3}{*}{\makecell[c]{Defense schemes}}} & \multicolumn{8}{c|}{Attacks}                                  & \multicolumn{1}{c}{\multirow{3}{*}{Papers}} \\
  \cline{3-10}    \multicolumn{2}{c|}{} & \multicolumn{2}{c|}{Physical Layer} & \multicolumn{2}{c|}{Preprocessing Layer} & \multicolumn{2}{c|}{Kernel Layer} & \multicolumn{2}{c|}{Service Layer} &  \\
  \cline{3-10}    \multicolumn{2}{c|}{} & TA    & VsA   & VsA   & AA    & SpA   & AA    & SqA   & FT    &  \\
      \hline
      \multicolumn{1}{c|}{\multirow{3}{*}{\makecell[c]{Liveness \\ detection}}} & \multicolumn{1}{c|}{\makecell[c]{Speaker \\ characteristics}} & $\bigotimes$      & $\bigoplus$      & $\bigoplus$      & $\bigoplus$      & $\bigoplus$      & $\bigoplus$      & $\bigotimes$      & $\bigotimes$      & \cite{ahmed2020void, blue2018hello, gong2018protecting, pradhan2019combating, wang2020differences, chen2017you} \\
  \cline{2-11}          & \multicolumn{1}{c|}{\makecell[c]{Human voice \\ characteristics}} & $\bigoplus$      & $\bigoplus$      & $\bigoplus$      & $\bigoplus$      & $\bigoplus$      & $\bigoplus$      & $\bigotimes$      & $\bigotimes$      & \cite{mochizuki2018voice, pradhan2019combating, shiota2016voice, wang2019secure, zhou2019hidden, meng2018wivo, shang2020voice, zhang2017hearing, feng2017continuous, sahidullah2017robust, shang2018defending, shang2019enabling, shang2020secure, zhang2020viblive, yang2023voshield} \\
  \cline{2-11}          & \multicolumn{1}{c|}{\makecell[c]{Active \\ interaction}} & $\bigoplus$      & $\bigoplus$      & $\bigoplus$      & $\bigoplus$      & $\bigoplus$      & $\bigoplus$      & $\bigodot$     & $\bigodot$     & \cite{carlini2016hidden, sugawara2020light, yuan2018all, rajaratnam2018isolated} \\
      \hline
      \multicolumn{1}{c|}{\multirow{3}{*}{\makecell[c]{Audio \\ conversion}}} & Encoding & $\bigotimes$      & $\bigodot$    & $\bigoplus$      & $\bigoplus$      & $\bigotimes$      & $\bigoplus$      & $\bigotimes$      & $\bigotimes$      & \cite{andronic2020mp3, das2019adagio, rajaratnam2018speech, zhang2019defending} \\
  \cline{2-11}          & Filtering & $\bigotimes$      & $\bigodot$     & $\bigoplus$     & $\bigoplus$      & $\bigotimes$      & $\bigoplus$      & $\bigotimes$      & $\bigotimes$      & \cite{chen2020devil, du2020sirenattack, yang2018characterizing, tamura2019novel, guo2020inor} \\
  \cline{2-11}          & Downsampling & $\bigotimes$      & $\bigodot$     & $\bigoplus$      & $\bigoplus$      & $\bigotimes$      & $\bigoplus$      & $\bigotimes$      & $\bigotimes$      & \cite{carlini2016hidden, chen2020devil, du2020sirenattack, yang2018characterizing, yuan2018commandersong, tamura2019novel} \\
      \hline
      \end{tabular}%
    \label{tab:general_compare}%
  \end{table*}%
  
  \begin{tcolorbox}
 Remark 4: As delineated in Table \ref{tab:general_compare}, liveness detection effectively counters a wide array of attacks, particularly those outside the service layer. While active interaction also offers some resistance against service layer attacks, its effectiveness comes at the cost of user experience. Given this trade-off, a preferable approach would be to combine passive detection with an enhanced service release and authentication mechanism. This combination would enable a more holistic defense against the diverse types of attacks encountered in VCS, optimizing both security and user experience.
\end{tcolorbox}

\section{Challenges and Future Directions}

\subsection{Challenges}
\emph{Hardware Enhancement.} The effectiveness of physical layer attacks in VCS often hinges on exploiting hardware vulnerabilities, particularly in microphones. However, these vulnerabilities are not uniformly present across all microphone types. A notable example is the iPhone 6 Plus, which has been shown to effectively resist voice-synthesis attacks due to its unique microphone design, as highlighted in \cite{he2019canceling}. This variability in hardware susceptibility presents significant challenges for executing consistent attacks at the physical layer.

\emph{Model Knowledge.} As VCS technology has evolved, commercial models have become increasingly prevalent. These models are often proprietary and not open-sourced, a strategy employed by companies to protect their intellectual property and prevent replication by competitors. This secrecy forces attackers to operate in a black-box environment, substantially lowering the success rate of adversarial attacks. Additionally, the challenge of creating universal adversarial perturbations that can produce similar attack outcomes across different models remains a significant hurdle in the realm of adversarial attacks.

\emph{Noise Disturbance.} Noise disturbance is a critical factor for both attackers and defenders in real-world VCS applications. For attackers, ambient noise can diminish the effectiveness and reach of malicious audio. Conversely, for defenders, noise can interfere with the accuracy of defense mechanisms, such as liveness detection systems, as evidenced in \cite{meng2022your}. Thus, both parties must account for the impact of noise in their strategies, which adds another layer of complexity to the security landscape of VCS.

\subsection{Future Directions}
The challenges identified in VCS security research open avenues for future work. To enhance the robustness of both attacks and defenses in practical scenarios, we propose the following directions:

\begin{itemize}
    \item \textbf{Focus on Black-box Attack Scenarios:} With the rise of closed-source models in VCS, attackers in real-world scenarios should concentrate on executing successful black-box attacks. This could involve crafting adversarial perturbations specifically for black-box models or designing transferable adversarial perturbations that can be applied from a known white-box model to an unknown black-box model.

    \item \textbf{Targeting Combined ASR and SV Functions:} Modern VCS, such as Apple's Siri, often integrate both Automatic Speech Recognition (ASR) and Speaker Verification (SV) functionalities. Attackers, therefore, need to develop malicious audio capable of simultaneously compromising both ASR and SV systems.

    \item \textbf{Optimizing Defense Schemes:} Future defense strategies should aim for high effectiveness while minimizing additional hardware requirements and system complexity. Leveraging existing device hardware for defense, as demonstrated in studies like \cite{ahmed2020void, meng2022your}, could offer more practical and efficient solutions.

    \item \textbf{Establishing Unified Standards for Evaluation:} The development of a unified standard for assessing attacks and defense mechanisms in VCS is crucial. This standard would provide reliable and consistent evaluation metrics, akin to the Common Vulnerability Scoring System (CVSS), aiding VCS designers in accurately assessing the security landscape.
\end{itemize}

These proposed directions aim to address the evolving landscape of VCS security, encouraging a balanced approach to unveiling sophisticated attacks and designing robust defenses.

\section{Conclusion}
This paper offers a thorough and methodical examination of the security landscape in VCS. We introduce a hierarchical model structure, dividing VCS into four layers: physical, preprocessing, kernel, and service. Within this framework, we develop a threat model and meticulously analyze the security threats unique to each layer, including transduction, voice-synthesis, adversarial, spoofing, squatting attacks, and false termination. We not only identify and describe the characteristics of these threats but also assess existing defense mechanisms and propose effective defense strategy combinations to enhance VCS security. Additionally, the paper provides strategic recommendations for future research, underscoring the necessity for ongoing innovation in addressing the evolving security challenges in VCS.

\bibliographystyle{ACM-Reference-Format}
\bibliography{ref}

\end{document}